# SCOTGRID: A PROTOTYPE TIER 2 CENTRE

A. Earl, P. Clark, S. Thorn*, University of Edinburgh, Edinburgh, Scotland

*Abstract*

ScotGrid [1] is a prototype regional computing centre formed as a collaboration between the universities of Durham, Edinburgh and Glasgow as part of the UK's national particle physics grid, GridPP [2]. We outline the resources available at the three core sites and our optimisation efforts for our user communities. We discuss the work which has been conducted in extending the centre to embrace new projects both from particle physics and new user communities and explain our methodology for doing this.

## INTRODUCTION

In this paper we present a summary of the Large Hadron Collider (LHC) computing infrastructure and ScotGrid's contribution and place within it. We describe the physical resources ScotGrid currently has available and the performance testing we have conducted on the Edinburgh high performance system for optimising users software. We present an account of our experiences with the LHC Computing Grid in terms of the system administration issues faced in joining the testbed and a user's view of performing analysis using it. We conclude with the future plans for the project.

### LHC Computing Grid

The CERN LHC is expected to start taking data in 2007, at which point the particle physics community will have to be capable of storing, analysing and cataloguing tens of petabytes of data per year. The LHC Computing Grid (LCG) aims to provide these capabilities in terms of simulation production, data analysis and data storage. This will be achieved through the use of a five tier hierarchical model, which makes use of the Grid computing paradigm.

This model comprises the four detectors associated with the LHC at Tier 0. National computing centres, which provide reliable data repositories and computing facilities to users within their country, and potentially internationally, are classed as Tier 1 sites. Regional computing centres, Tier 2, provide resources for a collection of institutes at a higher level of reliability and support than they may have locally. Individual institutes, such as a group within a university or research laboratory, are designated Tier 3 sites. These are not expected to support other groups and may not have significant data analysis or simulation production

---

* {aearl,pclark3,sthorn}@ph.ed.ac.uk

capabilities of their own. Within all sites, individual desktop computers are defined as Tier 4.

### GridPP

The GridPP collaboration [2] is the UK's contribution to the computing needs of the experiments based at the LHC: ATLAS, ALICE, CMS and LHCb. GridPP comprises a national (Tier 1) centre at Rutherford Appleton Laboratory (RAL) and four Tier 2 centres based around the geographical regions shown in Figure 1. These support 19 institutions and the UK particle physics community. GridPP also supports the UK's commitments to active particle physics experiments in the United States, including BaBar and D0.

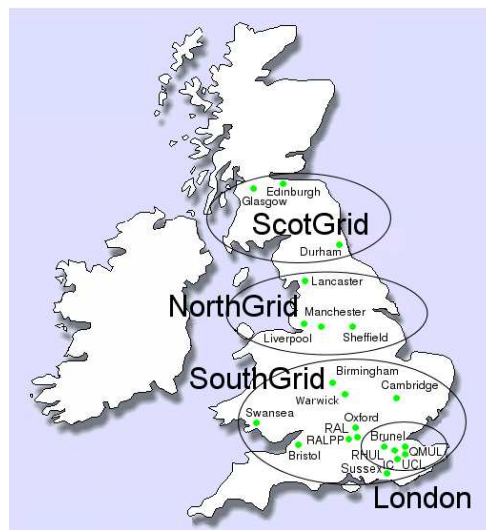

Figure 1: The GridPP collaboration

### The ScotGrid Tier 2 Centre

The ScotGrid prototype Tier 2 centre was started in 2001 as a collaboration between the Universities of Edinburgh and Glasgow to provide a distributed regional computing centre as part of the United Kingdom's GridPP project. The University of Durham later joined the project. The primary aim of this centre is to provide data analysis and simulation production facilities for the ATLAS and LHCb experiments. Further aims include supporting the particle physics community at the host institutes and others in the same geographical area.

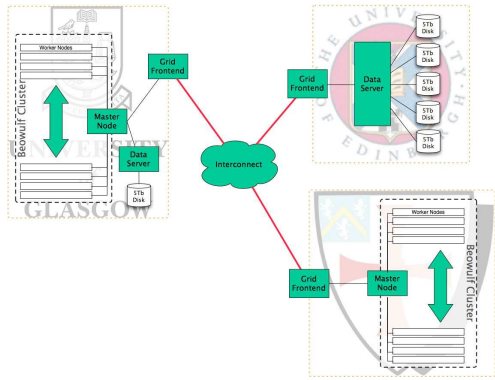

Figure 2: Logical layout of the ScotGrid system

*ScotGrid Resources*

In this section we provide a summary of the core resources available at the ScotGrid prototype Tier 2 centre. Further resources are available but are not included as part of the centre's core funding.

**Durham**

- Worker Nodes:
  - 40 × dual Intel P-IV 2.2GHz, 2GB RAM, 30GB disk

**Edinburgh**

- LCG front end systems:
  - 2 × IBM eServer x205 Intel P-IV 1.8GHz, 256MB RAM
  - IBM eServer x340 dual Intel P-III 1.0GHz, 2GB RAM
- Data Storage:
  - IBM eServer x440 8 Intel Xeon 1.9GHz, 32GB RAM
  - IBM Dual FAStT900 22TB RAID array
  - 10TB Sun Microsystems Storage Area Network
- Worker Nodes:
  - 4 × dual Intel Xeon 2.8GHz, 2GB RAM, 200GB EIDE HDD

**Glasgow**

- LCG front end systems:
  - 5 × Transtec 1U Intel PIV 2.6 GHz, 512Mb RAM
- Data Storage:
  - IBM eServer x370 Intel P-III Xeon 700MHz, 16GB RAM
  - IBM FAStT500 5TB RAID array
- Worker Nodes:
  - 5 × IBM eServer x340 dual Intel P-III 1GHz, 2GB RAM
  - 59 × IBM eServer x330 dual Intel P-III 1 GHz, 2GB RAM
  - 34 × IBM dual Intel Xeon 2.4GHz, 1.5GB RAM
  - 11 × Dell dual Intel Xeon 2.4GHz, 4GB RAM
  - 28 × IBM HS20 Blades dual Intel Xeon 2.4GHz, 1.5GB RAM
  - 6 × IBM x335 dual Intel Xeon 2.6GHz, 1.5GB RAM

## SYSTEM OPTIMISATION

The system in use at Edinburgh is an IBM x440 with eight Intel Xeon MP 1.90GHz processors and 32GB of memory. This gives us a theoretical maximum of 15.2GFlops based on a single floating point unit and one operation per cycle. We made use of the High Performance Linpack (HPL) benchmark [5] which was built using the Automatically Tuned Linear Algebra Software (ATLAS) libraries [6] to find the optimal performance. Both were compiled using the MPICH 1.2.6 C++ compiler. Other libraries, such as the Goto High Performance BLAS [7] were considered but not used due to time restrictions. For all tests we used a 2×4 processor arrangement.

The purpose of these tests is to find the optimal parameters for the system. We conducted two experiments, the first to discover the optimal matrix size and the second to discover the optimal block size at reasonable matrix size, using the results of the first test. In Figure 3 we show the results of the first experiment which varied the matrix size between 100 and 45,000 while using memory blocks of 100 to 500.

These results clearly show that the system does not reach its potential below a matrix size of 20,000 and that after this point performance has a greater reliance on block size than matrix size. In the early stages of the graph there is an inverse relationship between block size and performance i.e. small block sizes perform significantly better than larger ones with small problems. Conversely, for large problems a larger block size offers better performance than smaller ones.

The first experiment clearly showed that with a matrix size between 20,000 and 45,000 the major factor effecting performance was the size of the memory blocks used. For the second experiment we therefore conducted a more thorough test using a matrix of size 35,000 with a wider range of block sizes. This matrix size was chosen as it provides us with the size required for good performance, but with a lower total processing time than the maximum achievable.

As Figure 4 shows, increasing the block size between 80 and 320 improves performance significantly. After this point performance remains in the 10 - 11 GFlops range (65%-75% efficency) and we start to see diminishing returns above a block size of 440. A complete version of the results is available at http://www.ph.ed.ac.uk/ aearl/hpl/.

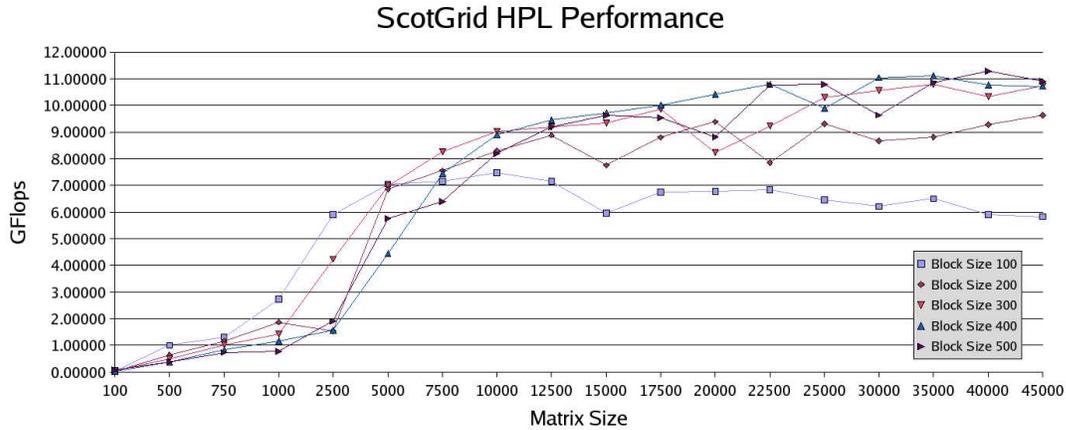

Figure 3: HPL results with varying block and matrix size

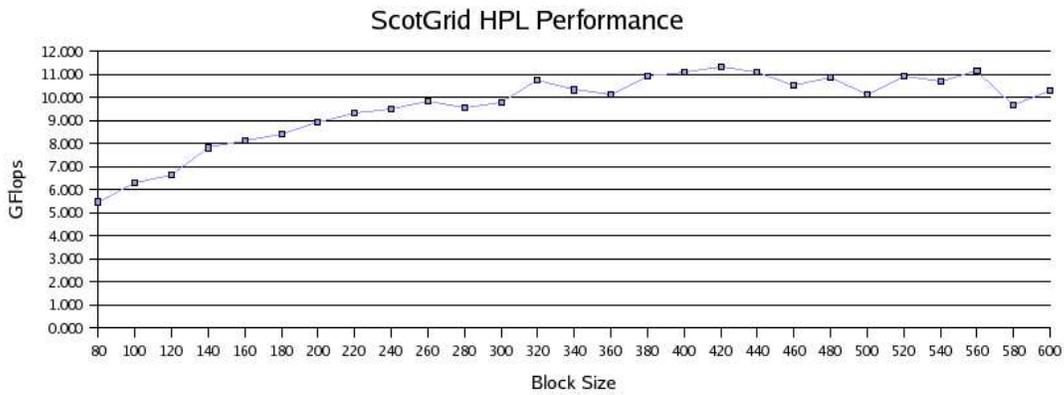

Figure 4: HPL results from varying block size

## DEVELOPING SCOTGRID

The obvious way to make the ScotGrid system available to the largest number of users, while minimising system administrator overhead, is to use standards based software. Where well defined standards are not available we make use of the most commonly used and supported software.

### A System Administrator's Experience of LCG

This section describes the experience of installing and configuring LCG software. Particular emphasis is placed on problems encountered and their resolution. Conclusions are drawn and an alternative strategy is proposed that should allow for faster deployment and knowledge transfer.

LCG was deployed using the LCFGng [8] installation and management tool. LCFGng installation was straightforward, however security issues with the use of what is currently an unsupported operating system, Red Hat Linux 7.3, was an extra overhead and concern.

Configuration of LCFGng, to allow installation of the LCG nodes, was found to be somewhat involved and not possible without assistance from the ScotGrid Tier 2 Coordinator and Tier 1 expertise.

Many problems arose from the storage biased hardware arrangement at Edinburgh. The original ScotGrid strategy was to concentrate on storage at the expense of processing power. This resulted in only a small number of individual machines that was insufficient for even a basic LCG installation. Furthermore, the switch providing connectivity between the equipment, a Cisco Catalyst 3550 12G, is a transceiver-based switch and consequently did not have any immediately available ports allowing quick addition of machines to resolve the shortfall. As a temporary solution, a fibre network card was purchased to allow a single/nominal Worker Node to be connected to the network at a different point.

Installation of the Computing Element (CE) node was relatively straightforward, the only issue being the support of two Ethernet interfaces. The Storage Element was rather difficult to install due to its hardware configuration, which includes three (different) network interfaces (two used), SCSI controllers and SCSI disk and a RAID controller. The network card providing external connectivity did not support PXE booting so a floppy disk was required including a custom kernel to support the hardware. The Worker Node (WN) installation was straightforward with a boot floppy.

However, much time was lost diagnosing the new network card that occasionally corrupted packets. This was solved with the replacement of the card.

Firewall issues were limited to the network's router, which performs basic packet filtering. The administration is handled by the University's Computing Service and is consequently out of direct control of the ScotGrid administrators. Problems were restricted to the Network Time Protocol (NTP) and Globus port range. Fortunately, the router acts as a stratum three time-server and so it was used instead of the GridPP time-servers (stratum two). The Globus port range on the router was different to the LCG default and connectivity was restored once this range was determined and the configuration updated.

To conclude, a considerable amount of extra work was required to handle the hardware and inflexibility of having only a few machines. A more efficient approach might be to conceive a prototype consisting of uniform desktop PCs connected to the internet with no intervening firewall. LCG installation using LCFGng should progress more smoothly and knowledge of the install process and functioning of LCG would be gathered quickly. The gained expertise could be transferred to the installation and configuration of the LCG production hardware and even be used to aide the determination of the most suitable hardware setup. The working prototype could be kept to provide a testbed for deploying new configurations and as a comparison/benchmark in the event of problems.

### A User's Experience of LCG for LHCb

The Edinburgh particle physics (experiments) group is an active member of several CP violation experiments, including LHCb. For the purposes of this paper we were assisted by a physicist who is active in analysing Monte Carlo data from the LHCb experiment in a non-Grid environment. With the assistance of the e-Science members of the group this physicist was taken through the stages of applying for a Grid certificate from the UK e-Science CA, and registering with the LHCb Virtual Organisation.

To submit jobs to the LCG testbed a User Interface (UI) is required. This is considerably more complex than the Globus Client software bundles and requires installation as the root user. In our university environment it is unusual for researchers to have superuser access on our production network, as this is regarded as a security risk. A machine was attached to our test network using DHCP for its IP address and with a firewall configured to prevent inbound communications. Red Hat 7.3 was installed and the UI was installed from RPMs.

This process was not the easiest, even with a freshly installed machine. Installation through the LCFG server was comparatively simpler, however it is unlikely that our users would be able to migrate to the Grid by themselves. Further issues included the lack of LCG documentation and basic instructions for selecting analysis environments.

From the experience we conclude that while simulation and data management on the Grid is well supported, further work must be undertaken to provide a similar level of capability for analysis. The software installation is not ready for less experienced users and the documentation will have to be developed considerably before it is used by new researchers.

## CONCLUSIONS

In this paper we have presented the current status of the ScotGrid Tier 2 centre and the results of benchmarking the high performance system at Edinburgh. Our early experiences with the LCG software have been presented from both a system administrator's and user's view point. ScotGrid intends to continue expanding its facilities in the run up to the start of LHC and hopes to increase the number of active users on its system and the number of universities involved with the centre.

## ACKNOWLEDGEMENTS

The equipment used in the ScotGrid project was funded by the Scottish Higher Education Funding Council (SHEFC) [3]. We would also like to thank the UK Particle Physics and Astronomy Research Council (PPARC) for funding studentship PPA/S/E/2001/03338 and the GridPP collaboration. Thanks also to the LHCb members of the University of Edinburgh particle physics (experiments) group who assisted in this paper and the members of EPCC who have provided advice and support.